\documentclass[notitlepage]{revtex4-1}

\usepackage[utf8]{inputenc}
\usepackage[colorlinks=true,citecolor=blue,linkcolor=blue]{hyperref}
\usepackage[normalem]{ulem}
\usepackage{url}
\usepackage{graphicx,wrapfig,float,slashed,cancel}
\usepackage{amsmath,amssymb,epsfig,graphicx,xcolor,stmaryrd}
\usepackage{bm}
\usepackage{enumitem}
\usepackage{multirow}

\definecolor{darkblue}{RGB}{1, 90, 173}

\begin{document}


\title{Investigation of $\Lambda(1405)$  as a molecular pentaquark state}

\author{K.~Azizi}
\email{ kazem.azizi@ut.ac.ir}
\thanks{Corresponding author}
\affiliation{Department of Physics, University of Tehran, North Karegar Avenue, Tehran
14395-547, Iran}
\affiliation{Department of Physics, Do\v{g}u\c{s} University, Dudullu-\"{U}mraniye, 34775
Istanbul, Turkey}
\affiliation{School of Particles and Accelerators, Institute for Research in Fundamental Sciences (IPM) P.O. Box 19395-5531, Tehran, Iran}
\author{Y.~Sarac}
\email{yasemin.sarac@atilim.edu.tr}
\affiliation{Electrical and Electronics Engineering Department,
Atilim University, 06836 Ankara, Turkey}
\author{H.~Sundu}
\email{ hayriyesundu.pamuk@medeniyet.edu.tr}
\affiliation{Department of Physics Engineering, Istanbul Medeniyet University, 34700 Istanbul, Turkey}

\date{\today}

\preprint{}

\begin{abstract}

$\Lambda(1405)$ is one of the interesting particles with its unclear structure and distinct properties. It has a  light mass compared to its non-strange counterpart, despite the strange quark it carries. This situation puts the investigation of this resonance among the hot topics in hadron physics and collects attention to clarify its properties. In this study, we focus on the calculation of the mass and residue of the $\Lambda(1405)$ resonance within the framework of QCD sum rules. We assign a structure in the form of a molecular pentaquark composed  from admixture of  $K^-$ meson-proton and  $\bar{K}^0$ meson-neutron. Using an interpolating current in this form, the masses and the current coupling constant are attained as $m=1406\pm 128~\mathrm{MeV}$ and $\lambda=(3.35\pm 0.35)\times10^{-5}~\mathrm{GeV}^6$ for $\slashed{q}$ and $m=1402\pm 141~\mathrm{MeV}$ and $\lambda=(4.08\pm 1.08)\times10^{-5}~\mathrm{GeV}^6$ for $I$ Lorentz structures entering the calculations, respectively. The obtained  mass values agree well with the experimental data supporting the plausibility of the considered structure.

\end{abstract}


\maketitle

\renewcommand{\thefootnote}{\#\arabic{footnote}}
\setcounter{footnote}{0}
\section{\label{sec:level1}Introduction}\label{intro} 

Among the well-know hyperons, the $\Lambda(1405)$ has an interesting place due to its peculiar properties, which are not easy to explain. This particle was first predicted theoretically in 1959~\cite{Dalitz:1959dn}, and its experimental verification came in 1961~\cite{Alston:1961zzd,Bastien:1961zz}. After that, there came many experimental reports for this state, some of which are given in the following Refs.~\cite{Thomas:1973uh,Braun:1977wd,Hemingway:1984pz,Ahn:2003mv,Zychor:2007gf,Niiyama:2008rt,HADES:2012csk,CLAS:2013rjt,CLAS:2013zie,CLAS:2014tbc,BGOOD:2021sog,J-PARCE31:2022plu}. In the quark mode, this state was predicted to be a first orbital excited state with $uds$ 3-quark content. Although this particle carries a strange quark, its unexpectedly lower mass than its corresponding non-strange counterparts makes it difficult to be explained by the quark model with a three-quark structure and to place it in the family of traditional baryons composed of three quarks. A comparison between the nucleon sector indicates a mass gap of around 600 MeV between P-wave excitation, $N(1535)$, and ground state, however, the same gap in the $\Lambda$ sector  is smaller than half of that of the nucleon sector. Another apparent distinction that cannot be explained by the simple quark model occurs in the mass difference between spin-orbit partners, namely the mass difference between $\Lambda(1405)$ and $\Lambda(1520)$ compared to that of $N(1535)$ and $N(1520)$.

After its announcement, the $\Lambda(1405)$ was first investigated by the quark model, placing it into a three-quark baryon family. However, taking into account this internal composition, its mass could not be explained by the constituent quark model~\cite{Isgur:1978xj,Capstick:1986ter}. In ref.~\cite{Isgur:1978xj}, the mass was obtained as $m\approx 1600$~MeV by conventional quark model. In Ref.~\cite{Nemoto:2003ft}, a mass of about $ 1.7$~GeV was predicted by Lattice QCD,  which is higher than the observed one, indicating the exclusion of the three-quark substructure. This discrepancy has focused attention on other possible structures. Due to its proximity to the $\bar{K}N$ threshold, the state was suggested to be a $\bar{K}N$ molecular-type bound state at its first prediction by Dalitz and Tuan~\cite{Dalitz:1959dn}, which was made several years before the proposal of the quark model. After the inconsistency of the quark model's prediction for the mass of the state, the meson baryon molecular interpretation was considered in various investigations. In Ref.~\cite{Hyodo:2011ur}, dynamical coupled-channel model was used, and the results favored a dominant meson baryon component for the structure of the $\Lambda(1405)$ state. The meson-baryon nature was investigated in Ref.~\cite{Hyodo:2008xr} by the chiral unitary approach, and from the analyses of meson-baryon scatterings, a dominant meson-baryon molecule component was obtained for the $\Lambda(1405)$. The structure and nature of this state were considered in  Refs.~\cite{Hyodo:2007np,Roca:2008kr} from its number of colors, $N_c$, behavior in the chiral unitary approach. The chiral unitary model was also applied in Ref.~\cite{Sekihara:2008qk} to get the electromagnetic mean squared radii of the $\Lambda(1405)$ giving a size larger than that of the ground state of ordinary baryons. $\bar{K}N$ bound state structure was also considered in Ref.~\cite{Akaishi:2002bg,Akaishi:2010wt}. In Ref.~\cite{Oset:1997it}, the coupled-channel Lippmann Schwinger equation was used to reproduce the properties of the $\Lambda(1405)$ resonance. In Ref.~\cite{Nakamoto:2006br}, a possibility for a mixed state was suggested for the $\Lambda(1405)$ underlying its possibility to be either a $q^4\bar{q}$ state or its mixture with a $q^3$ state. The investigation using the MIT Bag model~\cite{Strottman:1979qu} predicted a mass of $1400$~MeV for a $\Lambda^*$ state, treating it as a five-quark state. Pentaquark structure was also taken into account in Ref.~\cite{Xu:2022sak} using the constituent quark model, and $\Lambda(1405)$ was interpreted to have a possible structure as a mixture of the P-wave $q^2s$ state and the ground $q^3s\bar{q}$ pentaquark state. The possibility of the $\Lambda(1405)$ being a pentaquark state with $J^P=\frac{1}{2}^-$ was discussed in the Ref.~\cite{Zhang:2004xt} using Jaffe and Wilczek’s diquark model. In Refs.~\cite{Kaiser:1995eg,Kaiser:1996js} $\Lambda(1405)$ was considered as aquasi-bound state of $\bar{N}K$. The recent measurements~\cite{Zychor:2007gf,CLAS:2013rjt,BGOOD:2021sog} indicated a distortion from a single pole, whose reason is not clear yet. Two-pole structure identified by Ref.~\cite{Oller:2000fj} initiated studies of pole structure of the $\Lambda(1405)$ region. Analyzing the data provided by the CLAS collaboration, a chiral unitary framework was applied in Refs.~\cite{Mai:2014xna,Roca:2013cca}, which suggested the existence of two poles. In the recent experimental investigations by BGOOD collaboration~\cite{BGOOD:2021sog}, ALICE collaboration~\cite{ALICE:2022yyh}, and GlueX collaboration~\cite{Wickramaarachchi:2022mhi}, the two-pole picture was supported. On the other hand, recent J-PARC data~\cite{J-PARCE31:2022plu} was described by a single pole, and the Ref.~\cite{Anisovich:2020lec} provided a comprehensive analysis of the experimental data  indicated that a single resonance is sufficient to describe the data, however, the possibility of a two-pole model is not definitively dismissed. For further discussions on this, see the recent Refs.~\cite{BaryonScatteringBaSc:2023zvt,BaryonScatteringBaSc:2023ori,Guo:2023wes,Lu:2022hwm} and the references therein. As it is evident, despite numerous investigations supporting this issue, the two-pole structure of the $\Lambda(1405)$ region remains not entirely certain. This implies that both the single pole and two-pole structures remain as possibilities. In the current study, we chose to focus on the potential single pole picture and have analyzed the $\Lambda(1405)$ state accordingly.  Based on the need to clarify its obscure nature, various properties of the $\Lambda(1405)$ state, such as its mass and decays, were investigated using different approaches, see, for instance, the Refs.~\cite{Oller:2000fj,Hyodo:2008ek,Hyodo:2015rnm,Jido:2003cb,Jido:2009jf,Jido:2010ag,An:2010wb,MartinezTorres:2010zv,An:2010tv,Takahashi:2010nj,Sekihara:2010uz,Sekihara:2013sma,Sekihara:2014ica,Menadue:2011pd,MartinezTorres:2012yi,Mai:2012dt,Mai:2014xna,Guo:2012vv,Revai:2012fx,Oller:2013zda,Nakamura:2013boa,Nakamura:2013qda,Menadue:2013xqa,Dote:2014ema,Ohnishi:2014iba,Ohnishi:2015iaq,Hall:2014uca,Hall:2014gqa,Hall:2016kou,Nam:2015yoa,Miyahara:2015bya,Miyahara:2015cja,Miyahara:2015uya,Miyahara:2015eyq,He:2015cca,Fernandez-Ramirez:2015fbq,Molina:2015uqp,Kamiya:2016jqc,Liu:2016wxq,Dong:2016auh,Kim:2017nxg} and the references therein. 

The $\Lambda(1405)$ state has been listed as a four-star state in PDG~\cite{Workman:2022ynf}. On the other hand, as is seen, despite many theoretical and experimental affords to explain the nature and substructure of the $\Lambda(1405)$ state, there remains still uncertainty about its nature, and there exists no consensus on its structure. The possible structures predicted by various methods up to now, such as the meson-baryon molecule, a compact four-quark state, or a hybrid baryon, are in need of more support to justify or refute these probable structures. Therefore, studying $\Lambda(1405)$ to understand its nature is necessary and also contributes to and improves our understanding of the QCD at low energy.  On the other hand, a recent experimental investigation measuring the invariant mass spectra of a set of $\pi^{\pm}\Sigma^{\mp}$, $\pi^0\Sigma^0$, and $\pi^-\Sigma^0$~\cite{J-PARCE31:2022plu} has implied the possibility of the $\Lambda(1405)$ state being a temporary bound state of the $K^-$ meson and the proton. By this motivation in the present work, we consider a possible pentaquark substructure for this state in the $K^{-}p$ and $\bar{K}^0n$  meson-baryon molecular form. To this end, we apply a successful method, namely the QCD sum rules~\cite{Shifman:1978bx,Shifman:1978by,Ioffe81}, which gives many predictions over the observables of the particles consistent with experimental findings. In Refs.~\cite{Liu:1984dp,Choe:1997wz,Kondo:2006xz,Nakamura:2008zzc,Azizi:2017xyx,Kisslinger:2009dr}, to investigate $\Lambda(1405)$ state, the QCD sum rule method was used with different local operator choices such as formed with the combination of three quarks and a quark-antiquark pair, $\pi^0\Sigma^0$ multiquark interpolating field, three-quarks, mixing of three-quark and five-quark, a hybrid one with three-quark-one gluon content, and mixed hybrid and normal three-quark. The considered method requires a choice of a proper interpolating current for the state, and in this work, we chose the one with $K^{-}p$ and $\bar{K}^{0}n$ meson-baryon molecular form. Applying the method, we predict the mass of the $\Lambda(1405)$ whose comparison with experimental observation sheds light on the nature of the state. Besides, we also obtain the corresponding current coupling constant, which serves as an input for the calculation of the form factors applied in decay width calculations. 

The outline of the present  work is as follows: In the following section, Sec.~\ref{II}, we provide the details of the QCD sum rule calculations for the mass and current coupling constant. Sec.~\ref{III}  presents  the numerical analyses of the obtained results. The last section is devoted to the summary and conclusion.

\section{The QCD sum rule for the $\Lambda(1405)$ state}\label{II}

One of the effective ways to clarify the structure of a given resonance is to account for the mass of the state, considering a proper structure for the state and its comparison with experimental observations. To this end, in this work, we consider the $\Lambda(1405)$ state in the pentaquark substructure in the molecular form composed of $K^{-}p$ and $\bar{K}^{0}n$ and calculate the corresponding  mass with a choice of the interpolating current in this form using the QCD sum rule method. In this method, one initiates the calculations for the mass using the following correlation function:
\begin{equation}
\Pi(q)=i\int d^{4}xe^{iq\cdot
x}\langle 0|\mathcal{T} \{\eta_{\Lambda}(x)\bar{\eta}_{\Lambda}(0)\}|0\rangle,
\label{eq:CorrFmassLam}
\end{equation}
where $\eta_{\Lambda}$ is the interpolating current of $\Lambda(1405)$ with the mentioned form and  $\mathcal{T}$ denotes the time ordering operator. The current for the $\Lambda(1405)$ state is chosen in the molecular pentaquark form composed of $K^{-}p$ and $\bar{K}^{0}n$, with consideration of the quantum numbers the particle has, and it is given as follows:  
\begin{eqnarray}
\eta_{\Lambda}&=&\frac{1}{\sqrt{2}}[(\eta_{K^{-}})(\eta_{p})-(\eta_{\bar{K}^{0}})(\eta_{n})]\label{Current},
\end{eqnarray}
where
\begin{eqnarray}
\eta_{K^{-}}&=& \bar{u}_d \gamma_5 s_{d}\label{CurrentK},
\end{eqnarray}
\begin{eqnarray}
\eta_{\bar{K}^{0}}&=& \bar{d}_d \gamma_5 s_{d}\label{CurrentKbar},
\end{eqnarray}
\begin{eqnarray}
\eta_{p}&=&\sum_{i=1}^{2}2\epsilon^{abc}[u^{T}_{a}C A_1^i d_b]A_2^i u_c,\label{CurrentP}
\end{eqnarray}
\begin{eqnarray}
\eta_{n}&=&\sum_{i=1}^{2}2\epsilon^{abc}[d^{T}_{a}C A_1^i u_b]A_2^i d_c,\label{CurrentP}
\end{eqnarray}
and $C$ is charge conjugation operator, $a,~b,~c,~d$ are color indices of the quark fields corresponding to $u,~d~$ and $s$ quarks. The current is given in terms of  $A_1^i$ and $A_2^i$, which are $A_1^1=I$, $A_1^2=A_2^1=\gamma_5$, and $A_2^2=\beta$, with $\beta$ being an arbitrary parameter that is fixed later from the analyses.

The calculations involve two parts, which are called the QCD and the physical sides. In the QCD representation, the results are obtained in terms of QCD parameters such as quark and gluon condensates, and in the physical part in terms of physical parameters such as the mass and current coupling  of the considered state. By matching these two results, the obtained QCD sum rule gives the physical parameters in terms of QCD degrees of freedom. The match of the results  is performed via  dispersion integrals using the quark-hadron duality assumption. To suppress the contributions of the higher states and continuum, the Borel transformation and continuum subtraction are applied.

On the hadronic side, a complete set of hadronic states carrying the same quantum numbers with the state of interest and the corresponding interpolating current is inserted into the correlation function, and this gives us the hadronic representation in terms of the mass and the current coupling constant as  
\begin{eqnarray}
\Pi^{\mathrm{Had}}(q)= \frac{\langle 0|\eta_{\Lambda}|\Lambda(q,s)\rangle \langle \Lambda(q,s)|\bar{\eta}_{\Lambda}|0\rangle}{m^2-q^2}+\cdots,
\label{eq:masshadronicside1}
\end{eqnarray}
after taking the four-integral over $x$. In this expression, the contribution of the lowest state with negative parity is given explicitly, and those of higher states and continuum are represented by  $\cdots$. The  one particle state with momentum $q$ and spin $s$ is represented by $|\Lambda(q,s)\rangle $. Note that, in principle, the positive parity state also contributes to the correlation function, but it is expected that the mass of this positive parity state is significantly higher than that of the negative parity state. To this end, we adjust the continuum threshold, $s_0$, so that the contribution from the positive parity state is shifted into the contributions coming from excited states and continuum. The next step is to write the matrix element $\langle 0|\eta_{\Lambda}|\Lambda(q,s)\rangle$ in terms of the current coupling constant $\lambda$ and the spinor $u_{\Lambda}(q,s)$ as
\begin{eqnarray}
\langle 0|\eta_{\Lambda}|\Lambda(q,s)\rangle &=& \lambda u_{\Lambda}(q,s).
\label{eq:matrixelement1}
\end{eqnarray}
When we use this matrix element inside the Eq.~(\ref{eq:masshadronicside1}) and perform the summation over the spin via
\begin{eqnarray}
\sum_s u_{\Lambda}(q,s)\bar{u}_{\Lambda}(q,s)=\not\!q+m,
\end{eqnarray}
we get the hadronic side of the calculation as
\begin{eqnarray}
\Pi^{\mathrm{Had}}(q)=\frac{\lambda^2(\not\!q+m)}{m^2-q^2}+\cdots,
\end{eqnarray}
which is subsequently obtained as 
\begin{eqnarray}
\tilde{\Pi}^{\mathrm{Had}}(q)=\lambda^2e^{-\frac{m^2}{M^2}}(\not\!q+m)+\cdots,
\end{eqnarray}
after the Borel transformation with respect to $-q^2$, where $M^2$ is the corresponding Borel parameter. The $\tilde{\Pi}^{\mathrm{Had}}(q)$ is used to represent the result of the correlator after the Borel transformation. 

As stated, the computation of the physical parameter within the QCD sum rule method requires another calculation called the QCD side, which is done through operator product expansion (OPE). On this side, the correlator given in Eq.~(\ref{eq:CorrFmassLam}) is calculated using the interpolating current given in terms of the quark fields explicitly. When the quark fields are contracted via Wick's theorem, the correlation function turns into an expression given in terms of the quark propagators, as in the following equation:
\begin{eqnarray}
\Pi^{\mathrm{QCD}}(q)&=&i\int d^4x e^{iqx} 4\epsilon_{abc}\epsilon_{a'b'c'}\frac{1}{2}\mathrm{Tr}[S_u^{d'd}(-x)\gamma_{5}S_s{}^{dd'}(x)\gamma_{5}]\Big\{-\mathrm{Tr}[S_d^{bb'}(x)CS_u^T{}^{aa'}(x)C]\gamma_{5} S_{u}^{cc'}(x)\gamma_{5}\nonumber\\
&+& \gamma_{5}S_u^{ca'}(x)CS_d^T{}^{bb'}(x)CS_{u}^{ac'}(x)\gamma_{5}-\beta \mathrm{Tr}[S_d^{bb'}(x)\gamma_5 C S_u^T{}^{aa'}(x)C]\gamma_{5} S_{u}^{cc'}(x)\nonumber\\
&+& \beta \gamma_{5}S_u^{ca'}(x)\gamma_5 CS_d^T{}^{bb'}(x)C S_{u}^{ac'}(x)-\beta \mathrm{Tr}[S_d^{bb'}(x) C S_u^T{}^{aa'}(x)C\gamma_5] S_{u}^{cc'}(x)\gamma_{5}\nonumber\\
&+& \beta S_u^{ca'}(x) CS_d^T{}^{bb'}(x)C \gamma_5 S_{u}^{ac'}(x)\gamma_5-\beta^2 \mathrm{Tr}[S_d^{bb'}(x) \gamma_5 C S_u^T{}^{aa'}(x)C\gamma_5] S_{u}^{cc'}(x)\nonumber\\
&+&\beta^2 S_u^{ca'}(x)\gamma_5 CS_d^T{}^{bb'}(x)C \gamma_5 S_{u}^{ac'}(x)
\Big\}+\Big\{\substack{ u\rightarrow d \\ d\rightarrow u}\Big\}.
\label{Eq:PiQCDmass}
\end{eqnarray}   
To carry out the calculations, we need to use the following quark propagator inside the Eq.~(\ref{Eq:PiQCDmass}) explicitly~\cite{Reinders:1984sr,Yang:1993bp}
\begin{eqnarray}
S_{q,}{}_{ab}(x)&=&i\delta _{ab}\frac{\slashed x}{2\pi ^{2}x^{4}}-\delta _{ab}%
\frac{m_{q}}{4\pi ^{2}x^{2}}-\delta _{ab}\frac{\langle \overline{q}q\rangle
}{12} +i\delta _{ab}\frac{\slashed xm_{q}\langle \overline{q}q\rangle }{48}%
-\delta _{ab}\frac{x^{2}}{192}\langle \overline{q}g_{\mathrm{s}}\sigma
Gq\rangle +i\delta _{ab}\frac{x^{2}\slashed xm_{q}}{1152}\langle \overline{q}%
g_{\mathrm{s}}\sigma Gq\rangle  \notag \\
&&-i\frac{g_{\mathrm{s}}G_{ab}^{\alpha \beta }}{32\pi ^{2}x^{2}}\left[ %
\slashed x{\sigma _{\alpha \beta }+\sigma _{\alpha \beta }}\slashed x\right]
-i\delta _{ab}\frac{x^{2}\slashed xg_{\mathrm{s}}^{2}\langle \overline{q}%
q\rangle ^{2}}{7776}+\cdots ,  \label{Eq:qprop}
\end{eqnarray}%
where subindex $q$ is used to represent $u,~d~$ or  $s$ quark,  $a,~b=1,~2,~3$ are the color indices, and  $G^{\alpha\beta}_{ab}=G^{\alpha\beta}_{A}t_{ab}^{A}$. For the higher-order operators entering the calculation, the vacuum saturation approximation is employed. After the computation of the four-integral over $x$, the results are obtained as coefficients of two Lorentz structures, namely $\not\!q$ and $I$. In principle, any of these structures can be used for the QCD sum rule calculation. In this work, we obtain the results from both structures. Considering these structures, we gather the coefficients of the same structures from the hadronic and QCD sides and match them via a dispersion relation. To suppress the contributions coming from higher resonances and continuum and provide better convergence on the OPE side, we apply the Borel transformation to the results obtained on both sides. The match of the results of both sides gives
\begin{eqnarray}
\lambda^2 e^{-\frac{m^2}{M^2}}=\tilde{\Pi}^{\mathrm{QCD}}_{\slashed{q}}(s_0,M^2),\nonumber\\
\lambda^2 m e^{-\frac{m^2}{M^2}}=\tilde{\Pi}^{\mathrm{QCD}}_{I}(s_0,M^2),\label{QCDsumrule}
\label{QCDsumrule}
\end{eqnarray}
where $\tilde{\Pi}^{\mathrm{QCD}}_{\slashed{q}}(s_0,M^2)$ and  $\tilde{\Pi}^{\mathrm{QCD}}_{I}(s_0,M^2)$ represent the Borel transformed results obtained for the QCD side for $\slashed{q}$ and $I$ structures, respectively.  $\tilde{\Pi}^{\mathrm{QCD}}_{\slashed{q}(I)}(s_0,M^2)$ has the form  
\begin{eqnarray}
\tilde{\Pi}^{\mathrm{QCD}}_{\slashed{q}(I)}(s_0,M^2)=\int^{s_0}_{(3m_u+m_d+m_s)^2}ds  e^{-\frac{s}{M^2}}\rho_{\slashed{q}(I)}(s)+\Gamma_{\slashed{q}(I)}(M^2),\label{QCDsumruleOPE}
\end{eqnarray}
where the spectral densities, $\rho_{\slashed{q}(I)}(s)$, and $\Gamma_{\slashed{q}(I)}(M^2)$ are lengthy functions obtained from the computation of the QCD side and $\rho_{\slashed{q}(I)}(s)=\frac{1}{\pi}\mathrm{Im}[\Pi^{\mathrm{QCD}}_{{\slashed{q}(I)}}]$.  As examples $\rho_{\slashed{q}}(s)$, and $\Gamma_{\slashed{q}}(M^2)$ results obtained for the structure $\slashed{q}$ are presented in the Appendix to exemplify the results of the QCD side.

To get the mass of the considered state from  Eq.~(\ref{QCDsumrule}), we get the derivative of both sides with respect to  $-\frac{1}{M^2}$ and divide the resultant equation by Eq.~(\ref{QCDsumrule}) itself:
\begin{eqnarray}
m^2=\frac{\frac{d}{d(-\frac{1}{M^2})}\tilde{\Pi}^{\mathrm{QCD}}_{\slashed{q}(I)}(s_0,M^2)}{\tilde{\Pi}^{\mathrm{QCD}}_{\slashed{q}(I)}(s_0,M^2)},
\end{eqnarray} 
where $\tilde{\Pi}^{\mathrm{QCD}}(s_0,M^2)$ represents the right-hand side of Eq.~(\ref{QCDsumrule}), namely the Borel transformed results obtained in the QCD side of the calculation. The threshold parameter $s_0$ participates in the calculation as a result of the  continuum subtraction that is applied using the quark-hadron duality assumption. With the obtained mass, the current coupling constant is attained using the relation
\begin{eqnarray}
\lambda^2=e^{\frac{m^2}{M^2}}\tilde{\Pi}^{\mathrm{QCD}}_{\slashed{q}(I)}(s_0,M^2).
\end{eqnarray} 
The expressions obtained for mass and current coupling constant are used in the next section to get the numeric values of these quantities.

\section{Numerical Analyses}\label{III}

In this section, we obtain the numerical values of the mass and current coupling constant calculated in the previous section. To get their numerical values, there are some input parameters to be used in the results. Some of these input parameters are provided in Table~\ref{tab:Inputs}. 
\begin{table}[h!]
\begin{tabular}{|c|c|}
\hline\hline
Parameters & Values \\ \hline\hline
$m_{u}$                                    & $2.16^{+0.49}_{-0.26}~\mathrm{MeV}$ \cite{Workman:2022ynf}\\
$m_{d}$                                    & $4.67^{+0.48}_{-0.17}~\mathrm{MeV}$ \cite{Workman:2022ynf}\\
$m_{s}$                                     & $93.4^{+8.6}_{-3.4}~\mathrm{MeV}$ \cite{Workman:2022ynf}\\
$\langle \bar{q}q \rangle (1\mbox{GeV})$    & $(-0.24\pm 0.01)^3$ $\mathrm{GeV}^3$ \cite{Belyaev:1982sa}  \\
$\langle \bar{s}s \rangle $                 & $0.8\langle \bar{q}q \rangle$ \cite{Belyaev:1982sa} \\
$m_{0}^2 $                                  & $(0.8\pm0.1)$ $\mathrm{GeV}^2$ \cite{Belyaev:1982sa}\\
$\langle \overline{q}g_s\sigma Gq\rangle$   & $m_{0}^2\langle \bar{q}q \rangle$ \\
$\langle \frac{\alpha_s}{\pi} G^2 \rangle $ & $(0.012\pm0.004)$ $~\mathrm{GeV}^4 $\cite{Belyaev:1982cd}\\
\hline\hline
\end{tabular}%
\caption{The input parameters used in the numerical analyses.}
\label{tab:Inputs}
\end{table} 
However, these do not comprise all the input parameters that we need. Besides, we need three more auxiliary parameters that enter the calculation due to the Borel transformation, continuum subtraction, and the interpolating current used, which are $M^2$, $s_0$, and $\beta$, respectively. To fix the working intervals of these auxiliary parameters, we follow some criteria that are standard for the QCD sum rule calculations. Among these criteria is the stability of the results with the variation of these parameters. Another criterion is the one used to fix the value of the threshold parameter. The threshold parameter is related to the energy of the possible lowest excited state of the considered one, above which we take the states as part of the continuum. Requiring the minimum contribution of these states, in other words, demanding a dominant contribution from the considered state compared to the continuum, we set a proper interval satisfying this condition from the analyses of the results. Our analyses lead to the interval given as
\begin{eqnarray}
&2.3~\mbox{GeV}^2 \leq s_0 \leq 2.9~\mbox{GeV}^2.&
\end{eqnarray} 

\begin{figure}[h!]
\begin{center}
\includegraphics[totalheight=5cm,width=7cm]{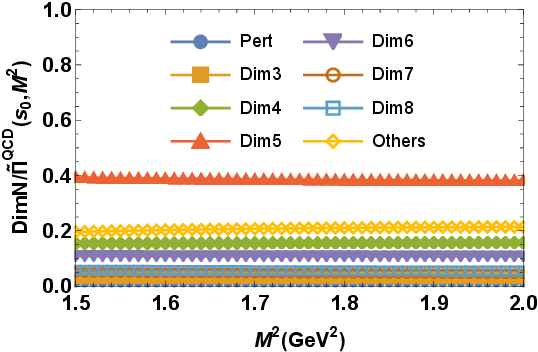}
\includegraphics[totalheight=5cm,width=7cm]{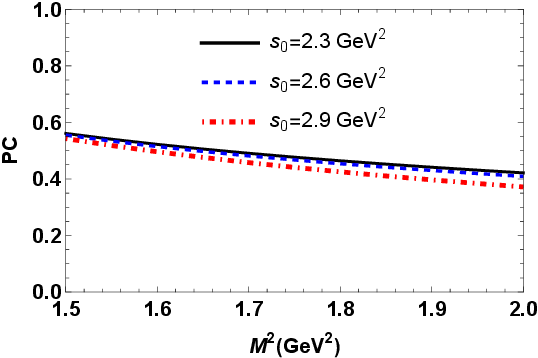}
\end{center}
\caption{\textbf{Left:} Contributions of different dimensions in the QCD side's result as a ratio of the contribution of each dimension to the total QCD result at the central value of $s_0$.
\textbf{Right:} The pole contribution for different values of $s_0$. }
\label{gr:dimension}
\end{figure}

For the interval of the Borel parameter $M^2$, the one that ensures the convergence of the OPE and dominance of pole   is chosen. To this, we seek a region where the contribution of the higher-order terms on the QCD side is small enough (sets the lower limit on  $M^2$) and the contribution of the lowest state is dominant over that of the higher states and continuum (sets the upper limit on  $M^2$).
To demonstrate the fulfillment of these requirements,  we present the contributions of different dimensions in the QCD side's result  as a ratio of the contribution of each dimension to the total QCD result in Figure~\ref{gr:dimension} considering the results obtained from the $\slashed{q}$ structure. We set the lower limit for  $M^2$ by analyzing the contributions of the higher dimensional terms in the OPE and the condition of the convergence of the obtained series.   In the same figure, we present the pole contribution (PC) as well. We set the upper limit for the working region of the Borel parameter, requiring the pole contribution to be at least or greater than  $35\% $, which is a pretty good value for the pentaquark states:
\begin{eqnarray}
PC(s_0,M^2)=\frac{\tilde{\Pi}^{QCD}(s_0,M^2)}{\tilde{\Pi}^{QCD}(\infty,M^2)}\gtrsim 35\% .
\end{eqnarray}
Via the analyses, we fix the window for $M^2$ as
\begin{eqnarray}
1.5~\mbox{GeV}^2\leq M^2\leq 2.0~\mbox{GeV}^2.
\end{eqnarray}

For the  interval of the final parameter, $\beta$, the results are analyzed via a parametric plot in which the variation of the results is considered as a function of $\cos \theta$, where $\beta=\tan \theta$. From the analyses, the regions with the relative least variation are chosen, and the obtained regions are as follows:
\begin{eqnarray}
-1.0 \leq \cos \theta \leq -0.5 ~~~~~~~ \mathrm{and} ~~~~~~~~0.5 \leq \cos \theta \leq 1.0.
\end{eqnarray} 

The input and auxiliary parameters are used in the QCD sum rule results for the mass and the current coupling constant to get their corresponding values. From the calculations, the following values are obtained:
\begin{eqnarray}
m=1406\pm 128~\mathrm{MeV}~~~~~~\mathrm{and}~~~~~~\lambda=(3.35\pm 0.35)\times10^{-5}~\mathrm{GeV}^6,
\end{eqnarray}
for the $\slashed{q}$ structure and 
\begin{eqnarray}
m=1402\pm 141 ~\mathrm{MeV}~~~~~~\mathrm{and}~~~~~~\lambda=(4.08\pm 1.08)\times10^{-5}~\mathrm{GeV}^6,
\end{eqnarray}
for the $I$ structure. As expected, the mass values derived from these two structures exhibit consistency with each other. The errors in the results enter the calculations due to the uncertainties of the input parameters and the determination of the working windows of  auxiliary parameters. 

\begin{figure}[h!]
\begin{center}
\includegraphics[totalheight=5cm,width=7cm]{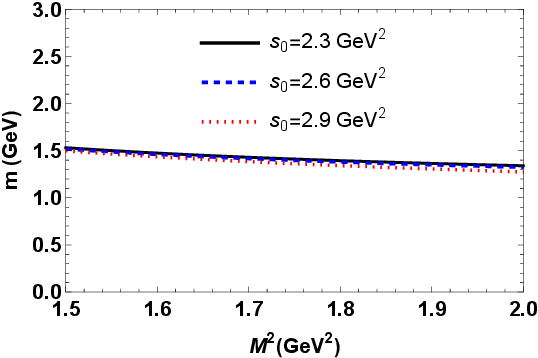}
\includegraphics[totalheight=5cm,width=7cm]{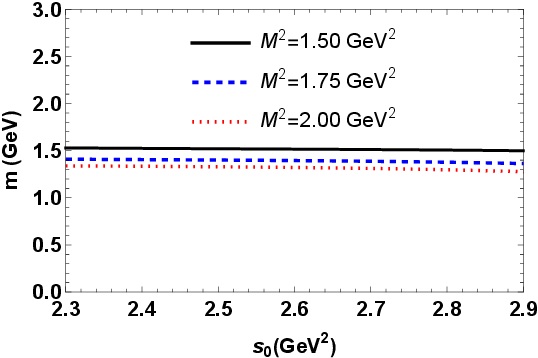}
\end{center}
\caption{\textbf{Left:} The dependence  of the mass on the Borel parameter $M^2$ at different values of the threshold parameter $s_0$.
\textbf{Right:} The dependence of the mass on the threshold parameter $s_0$ at different values of the  Borel parameter   $M^2$. }
\label{gr:MassMsqS0}
\end{figure}

\begin{figure}[h!]
\begin{center}
\includegraphics[totalheight=5cm,width=7cm]{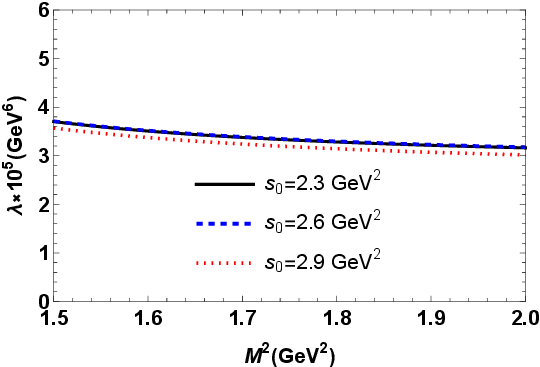}
\includegraphics[totalheight=5cm,width=7cm]{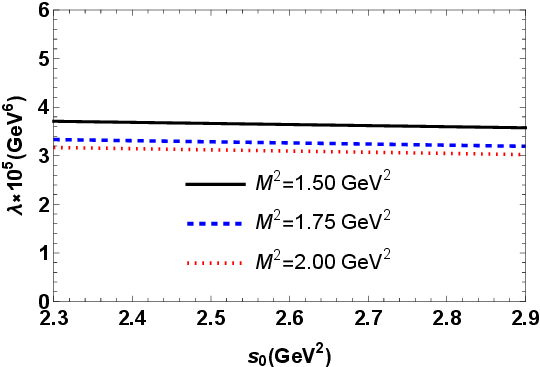}
\end{center}
\caption{\textbf{Left:} The dependence of the current coupling constant on the Borel parameter $M^2$ at different values of the threshold parameter $s_0$.
\textbf{Right:} The dependence of the current coupling constant on the threshold parameter $s_0$ at different values of the Borel parameter $M^2$. }
\label{gr:residueMsqS0}
\end{figure}

Finally, to show the variations of the obtained mass and the current coupling constant as a function of the auxiliary parameters in the chosen working intervals of the Borel parameter and continuum threshold, we plot the Figures~\ref{gr:MassMsqS0} and~\ref{gr:residueMsqS0} for the results obtained from  the $\slashed{q}$ structure. 
As the figures show, the mild variation requirement of the obtained results in these regions is satisfied as required.

\section{Summary and conclusion}\label{IV}

Since its first observation, the $\Lambda(1405)$ has been one of the intriguing states, with properties different from those expected by the quark model. It has attracted interest with its lightest mass among the negative parity baryons, in spite of the strange quark it carries. Furthermore, the distortion of its line shape from a Breit-Wigner form might be an indication of its exotic nature. 

In Ref.~\cite{Choe:1997wz}, this state was considered via the QCD sum rule method in a multiquark form with a $\pi^0\Sigma^0$ interpolating field, which resulted in a prediction of the mass of $1.419~\mathrm{GeV}$. In Ref.~\cite{Kondo:2006xz}, QCD sum rule method was applied to see if the $\Lambda(1405)$ is a parity partner of the ground state $\Lambda$ baryon, and the results based on the obtained mass predictions indicated this could not be the case. The mixing of three-quark and five-quark Fock components was taken into account in Ref~\cite{Nakamura:2008zzc}, and the QCD sum rule analyses resulted in the coupling to the five-quark operator being much stronger. The QCD sum rule method was also used with a hybrid state assumption in Ref~\cite{Azizi:2017xyx} with three-quark-one gluon content, and the corresponding mass was obtained as $m=1403^{+33}_{-32}~\mathrm{MeV}$. In that work, the $\Lambda(1405)$ was also investigated as $P$-wave ordinary three-quark state, the predicted mass was $m=1435^{+32}_{-31}~\mathrm{MeV}$, revealing that the hybrid structure has a better consistency with the experimental mass value~\cite{Azizi:2017xyx}. Another prediction with the QCD sum rule for the $\Lambda(1405)$ being a hybrid state was given in Ref~\cite{Kisslinger:2009dr}. All these works, performed with alternative structures but with predictions consistent with the experimental observation within the errors, indicate the need for more investigation for the $\Lambda(1405)$ state. To improve our  understanding of this  state, it was necessary to consider this state once more.  With this motivation, among the various suggested structures, in this work, we focused on the pentaquark one. We adopted an interpolating current in the molecular form composed of the combination of $K^-p$ and $\bar{K}^0n$ and predicted the corresponding mass of the state. We have extracted the mass as $m=1406\pm 128~\mathrm{MeV}$ for $\slashed{q}$ structure and $m=1402\pm 141 ~\mathrm{MeV}$ for $I$ structure, which indicate that combined $K^-p$ and $\bar{K}^0n$ pentaquark structure leads to a mass result well consistent with the experimentally observed one, suggesting that  this structure is the  most probable one for the considered state. It is worth noting that the present study differs from prior investigations that employed the QCD sum rules method in that the considered structure for the state are different. All these possible different structures considered in various studies and the present one have yielded mass predictions in accordance with experimental observations, which clearly implies the need for further investigations on other properties of the state with the objective of elucidating its substructure and reaching a conclusive determination. Such additional investigations may provide either supports or dismiss the proposed structure and help identify the structure that most accurately characterizes the state under consideration.

Additionally, we obtained the current coupling constant, which is  $\lambda=(3.35\pm 0.35)\times10^{-5}~\mathrm{GeV}^6$ for $\slashed{q}$  and $\lambda=(4.08\pm 1.08)\times10^{-5}~\mathrm{GeV}^6$ for $I$ structure. The current coupling constant is one of the main inputs that is required to calculate the physical quantities like different form factors, decay width, branching fraction, etc. in any weak, electromagnetic or  strong decays of  $\Lambda(1405)$. Therefore, it is of great importance to fix its value for further investigations of this state. The results obtained in this study may provide valuable insights to unravel the nature of this intriguing baryon resonance.

\section{\label{appendix}Appendix: QCD Side of the Correlation Function}

In this appendix, to exemplify the results obtained for the QCD side of the calculation, we choose the $\slashed{q}$ structure and present the result of $\tilde{\Pi}^{\mathrm{QCD}}_{\slashed{q}}(s_0,M^2)$ obtained after Borel transformation and continuum subtraction in the following form:
\begin{eqnarray}
\tilde{\Pi}^{\mathrm{QCD}}_{\slashed{q}}(s_0,M^2)=\int_{s_{0,\mathrm{min}}}^{s_{0}}ds\rho_{\slashed{q}}^{\mathrm{OPE}%
}(s)e^{-s/M^{2}}+\Gamma_{\slashed{q}}(M^2),
\end{eqnarray}
where  $s_{0,\mathrm{min}}=(3m_u+m_d+m_s)^2$ is the minimum energy required for the creation of the considered particle, $\Lambda(1405)$. The spectral density $\rho_{\slashed{q}}^{\mathrm{OPE}}(s)$  and $\Gamma_{\slashed{q}}(M^2)$ are given as
formula
\begin{eqnarray}
\rho_{\slashed{q}}^{\mathrm{OPE}}(s)=\rho ^{\mathrm{pert.}}(s)+\sum_{N=3}^{13}\rho ^{\mathrm{DimN}}(s),
\end{eqnarray}
and 
\begin{eqnarray}
\Gamma_{\slashed{q}}(M^2)=\sum_{N=9}^{27}\tilde{\Pi}^{DimN}.
\end{eqnarray}
The functions $\rho^{\mathrm{pert.}}(s )$, $ \rho^{\mathrm{DimN}}(s) $ and $\tilde{\Pi}^{DimN}$ are as follows:
\begin{eqnarray}
\rho ^{\mathrm{pert.}}(s )&=&\dfrac{s^{5}(5 t^{2}+2 t+5)}{7\times 5^{2}\times 3 \times 2^{20}  \pi^8} ,
\nonumber \\
\rho ^{\mathrm{Dim3}}(s )&=& \big(\langle \bar{s} s \rangle- 2 \langle \bar{d} d \rangle \big)\dfrac{m_s s^3 (5 t^{2}+2 t+5)}{5 \times 3\times 2^{16}  \pi^6} ,
\nonumber \\
\rho ^{\mathrm{Dim4}}(s )&=& \langle g_s^2 G^2\rangle \dfrac{ s^3 (5 t^{2}+2 t+5)}{5 \times  2^{20} \pi^8} ,
\nonumber \\
\rho ^{\mathrm{Dim5}}(s )&=&m_0^2\left[ 4 \langle \bar{s} s \rangle +\langle\bar{d} d\rangle  \left( 49+ ln( \frac{s}{\Lambda^2})\right) \right] \dfrac{ m_s s^2 (5 t^{2}+2 t+5)}{3 \times  2^{18}  \pi^6} ,
\nonumber \\
\rho ^{\mathrm{Dim6}}(s )&=&\langle \bar{u} u \rangle^{2}\dfrac{s^2\left[54  \pi^2 (t-1)^2+g_s^2 (5 t^{2}+2 t+5)\right] }{3^4\times 2^{14} \pi^6} + \left( 2 \langle \bar{d} d \rangle^{2}+\langle \bar{s} s \rangle^{2}\right) \dfrac{g_s^2 s^2 (5 t^{2}+2 t+5)}{3^4\times 2^{15} \pi^6} 
\nonumber \\
&+& \langle \bar{s} s \rangle \langle \bar{d} d \rangle \dfrac{ s^2 (5 t^{2}+2 t+5)}{3 \times 2^{13} \pi^4} +\langle \bar{u} u \rangle \langle \bar{d} d \rangle \dfrac{ s^2 (t^2-1)}{  2^{12}  \pi^4} ,
\nonumber \\
\rho ^{\mathrm{Dim7}}(s )&=& \langle g_s^2 G^2 \rangle \left[ 3  \langle \bar{s} s \rangle -  \langle \bar{d} d\rangle \left( 19- ln (\dfrac{s}{\Lambda^2})\right)  \right] \dfrac{m_s s (5 t^{2}+2 t+5)}{3^2 \times 2^{16}  \pi^6 } ,
\nonumber \\
\rho ^{\mathrm{Dim8}}(s )&=& \langle g_s^2 G^2 \rangle^{2}\frac{s (7 t^2+12 t+7)}{3^2 \times 2^{20}  \pi^8}- m_0^2 \langle \bar{u} u \rangle^2 \dfrac{s (t-1)^2}{3 \times 2^{12}  \pi^4}-m_0^2 \langle \bar{u} u \rangle \langle \bar{d} d \rangle  \dfrac{s (t^2-1)}{3 \times 2^{10}  \pi^4}-m_0^2 \langle \bar{s} s \rangle \langle \bar{d} d \rangle  \dfrac{s (5 t^2+2 t +5)}{3 \times 2^{11}  \pi^4},
\nonumber \\
\rho ^{\mathrm{Dim9}}(s )&=&m_0^2 \langle \bar{d} d \rangle \langle g_s^2 G^2 \rangle \frac{m_s (5 t^2 + 2 t + 5) }{ 3^{4}\times 2^{21} \pi^6}\left\lbrace 91+50 \gamma_{E}-50 ln( \dfrac{s}{\Lambda^2}) \right\rbrace - m_0^2 \langle \bar{s} s \rangle \langle g_s^2 G^2 \rangle \frac{7 m_s (5 t^2 + 2 t + 5) }{ 3\times 2^{21} \pi^6}
\nonumber \\
&+& g_s^2  \langle \bar{d} d \rangle^3 \frac{m_s (5 t^2 + 2 t + 5) }{ 3^{4}\times 2^{10} \pi^4}+\langle \bar{d} d \rangle \langle \bar{u} u \rangle \langle \bar{s} s \rangle \dfrac{m_s (t^2-1)}{2^7 \pi^2}-\langle \bar{d} d \rangle^2 \langle \bar{u} u \rangle\dfrac{m_s (t^2-1)}{2^6 \pi^2} 
\nonumber \\
&+&
\langle \bar{d} d \rangle^2 \langle \bar{s} s \rangle \dfrac{g_s^2 m_s(5t^2 + 2 t +5)}{3^4 2^{10}\pi^4} -\langle \bar{u} u \rangle^2 \langle \bar{d} d \rangle  \dfrac{m_s \left[ 54 \pi^2 (t-1)^2+g_s^2(5 t^2+2 t +5)\right] }{3^4 2^{9}\pi^4}
\nonumber \\
&+&\langle \bar{u} u \rangle^2 \langle \bar{s} s \rangle  \dfrac{m_s \left[ 54 \pi^2 (t-1)^2+g_s^2(5 t^2+2 t +5)\right] }{3^4 2^{10}\pi^4},
\nonumber \\
\rho ^{\mathrm{Dim10}}(s )&=&\langle \bar{d} d \rangle^2 \langle g_s^2 G^2 \rangle \dfrac{g_s^2 (7 t^2 + 2 t +7)}{3^{4}\times 2^{14} \pi^6}+\langle \bar{u} u \rangle^2 \langle g_s^2 G^2 \rangle \dfrac{\left[ 432 \pi^2 (t-1)^2+4 g_s^2 (13 t^2 + 6 t +13)\right] }{3^{4}\times 2^{16} \pi^6}
\nonumber \\
&+&
\langle \bar{s} s \rangle^2 \langle g_s^2 G^2 \rangle \dfrac{g_s^2(5 t^2+ 2 t +5)}{3^4\times 2^{15} \pi^6}+ m_0^4 \langle \bar{u} u \rangle \langle \bar{d} d \rangle \dfrac{3 (t^2-1)}{2^{13}\pi^4}+m_0^4 \langle \bar{s} s \rangle \langle \bar{d} d \rangle \dfrac{3 (5t^2+2 t+5)}{2^{14}\pi^4}
\nonumber \\
&+&\langle \bar{u} u \rangle \langle \bar{d} d \rangle \langle g_s^2 G^2 \rangle \dfrac{(t^2-1)}{2^{11} \pi^4}+\langle \bar{s} s \rangle \langle \bar{d} d \rangle \langle g_s^2 G^2 \rangle \dfrac{(5 t^2+2 t +5)}{3\times 2^{13} \pi^4},
\nonumber \\
\rho ^{\mathrm{Dim11}}(s )&=&\langle \bar{d} d \rangle \langle g_s^2 G^2 \rangle^2 \dfrac{m_s (5 t^2 +2 t +5)}{3\times 2^{17} M^2  \pi^{6}} ln( \dfrac{s}{\Lambda^2}) -m_0^2  \langle \bar{d} d \rangle^3 \dfrac{g_s^2 m_s (5 t^2 +2 t +5)}{3^{4}\times 2^{10} M^2 \pi^4} ln( \dfrac{s}{\Lambda^2}) 
\nonumber \\
&-&m_0^2 \langle \bar{u} u \rangle^2 \langle \bar{d} d \rangle  \dfrac{m_s\left[54 \pi^2  (t-1)^2 +g_s^2(5 t^2 + 2 t +5)\right] }{3^4\times 2^{9} M^2 \pi^4} ln( \dfrac{s}{\Lambda^2}) -m_0^2 \langle \bar{u} u \rangle \langle \bar{d} d \rangle ^2 \dfrac{m_s(t^2-1)}{2^7 M^2 \pi^2} ln( \dfrac{s}{\Lambda^2}) ,
\end{eqnarray}

\begin{eqnarray}
\rho ^{\mathrm{Dim12}}(s )&=&0,
\nonumber \\
\rho ^{\mathrm{Dim13}}(s )&=& m_0^2 \langle \bar{d} d \rangle \langle g_s^2 G^2 \rangle^2 \dfrac{m_s (5 t^2 +2 t +5)}{3\times 2^{19} M^4 \pi^6} ln ( \dfrac{s}{\Lambda^2}) +\langle \bar{d} d \rangle^3 \langle g_s^2 G^2 \rangle \dfrac{g_s^2 m_s (5 t^2 +2 t+ 5)}{3^5\times 2^{12} M^4 \pi^4} ln( \dfrac{s}{\Lambda^2})
\nonumber \\
&+&
 \langle \bar{d} d \rangle ^2 \langle \bar{u} u \rangle \langle g_s^2 G^2 \rangle \dfrac{m_s(t^2-1)}{3\times 2^{9}M^4 \pi^2} ln( \dfrac{s}{\Lambda^2})+ \langle \bar{d} d \rangle  \langle \bar{u} u \rangle ^2\langle g_s^2 G^2 \rangle \dfrac{m_s \left[ 54 \pi^2 (t-1)^2+g_s^2 (5 t^2 +2 t +5)\right] }{3^5 \times 2^{11}M^4 \pi^4}ln( \dfrac{s}{\Lambda^2})
 \nonumber \\
&+&
\left( m_0^4 \langle \bar{d} d \rangle \langle \bar{u} u \rangle^2 + 6 m_0^4 \langle \bar{d} d \rangle^2  \langle \bar{u} u \rangle \right)\dfrac{m_s (t^2-1)}{3\times 2^{9} M^4 \pi^2} ln( \dfrac{s}{\Lambda^2}),
 \nonumber \\
\tilde{\Pi}^{Dim9}&=&m_0^2 \langle \bar{d} d \rangle \langle g_s^2 G^2 \rangle \dfrac{25 \gamma_{E} m_s  M^2 (5 t^2 + 2 t + 5) }{3\times 2^{20} \pi^6} \left( e^{-\dfrac{s_0}{M^2}}-1\right) ,
\nonumber \\
\tilde{\Pi}^{Dim10}&=&0,\nonumber\\
\tilde{\Pi}^{Dim11}&=&m_0^2 \langle \bar{d} d \rangle^3 \dfrac{g_s^2 (3-2\gamma_{E})m_s (5 t^2+ 2 t +5)}{3^{4}\times 2^{11} \pi^4}+m_0^2 \langle \bar{u} u \rangle^2 \langle \bar{s} s \rangle\dfrac{m_s \left[ 27 \pi^2(t-1)^{2}+2 g_s^2(5 t^2 + 2 t + 5) \right] }{3^5\times 2^{10}\pi^4}
\nonumber \\
&-&m_0^2 \langle \bar{u} u \rangle^2 \langle \bar{d} d \rangle \dfrac{m_s\left[108 (\gamma_{E}-2) \pi^2 (t-1)^2+g_s^2  (2-3\gamma_{E})(5 t^2 + 2 t +5)\right] }{3^4 \times 2^{10} \pi^4}
\nonumber \\
&+&
\langle \bar{d} d \rangle  \langle g_s^2 G^2 \rangle^2\dfrac{(3\gamma_{E}-1)m_s (5 t^2+ 2 t +5) }{3^2\times 2^{17}\pi^6}-m_0^2 \langle \bar{u} u \rangle \langle \bar{d} d \rangle^2 \dfrac{m_s (2\gamma_{E}-5)(t^2-1)}{2^8 \pi^2}
\nonumber \\
&+&
m_0^2 \langle \bar{d} d \rangle^2\langle \bar{s} s \rangle \dfrac{g_s^2 m_s (5 t^2+2 t+5)}{3^5\times 2^{11} \pi^4} 
+m_0^2 \langle \bar{d} d \rangle^2\langle \bar{u} u \rangle \dfrac{ m_s (5 -2\gamma_{E})(t^2-1)}{ 2^{8} \pi^4}-m_0^2 \langle \bar{d} d \rangle \langle \bar{u} u \rangle\langle \bar{s} s \rangle \dfrac{ m_s (t^2 -1)}{ 3\times 2^{8} \pi^4},
\nonumber \\
\tilde{\Pi}^{Dim12}&=&\left( \langle \bar{d} d \rangle^4+\langle \bar{u} u \rangle^4 + \langle \bar{d} d \rangle^2 \langle \bar{s} s \rangle^2
\right)  \dfrac{g_s^4 (5 t^2 +2 t+5)}{3^8\times 2^9 \pi^4}+ \langle \bar{s} s \rangle \langle \bar{d} d \rangle \langle \bar{u} u \rangle^2 \dfrac{\left[54 \pi^2 (t-1)^2+g_s^2 (5 t^2 +2 t +5) \right] }{3^5 \times 2^6 \pi^2}
\nonumber \\
&+&
 \left( \langle \bar{s} s \rangle^2 \langle \bar{u} u \rangle^2 +2  \langle \bar{d} d \rangle^2 \langle \bar{u} u \rangle^2  \right) \dfrac{\left[ 54 g_s^2 \pi^2 (t-1)^2+g_s^4 (5 t^2+2 t+ 5)\right] }{3^8\times 2^8 \pi^4}+  \langle \bar{s} s \rangle \langle \bar{u} u \rangle \langle \bar{d} d \rangle^2 \dfrac{(t^2-1)}{3\times 2^4}
 \nonumber \\
&+&
\left( \langle \bar{u} u \rangle^3 \langle \bar{d} d \rangle+\langle \bar{u} u \rangle \langle \bar{d} d \rangle^3+ \langle \bar{u} u \rangle \langle \bar{d} d \rangle \langle \bar{s} s \rangle^2
\right)  \dfrac{g_s^2 (t^2-1)}{3^4 \times 2^6 \pi^2}+\langle \bar{d} d \rangle^3 \langle \bar{s} s \rangle \dfrac{g_s^2 (5 t^2+2 t +5)}{3^5 \times 2^7 \pi^2}
\nonumber \\
&-&
m_0^2 \langle \bar{d} d \rangle \langle \bar{s} s \rangle \langle g_s^2 G^2 \rangle \dfrac{(5 t^2+ 2 t +5)}{3\times 2^{13} \pi^4}-m_0^2 \langle \bar{d} d \rangle \langle \bar{u} u \rangle \langle g_s^2 G^2 \rangle \dfrac{(t^2-1)}{ 2^{11} \pi^4}-m_0^2 \langle \bar{u} u \rangle^2  \langle g_s^2 G^2 \rangle \dfrac{(t-1)^2}{3\times 2^{13}\pi^4},
\nonumber \\
\tilde{\Pi}^{Dim13}&=&- m_0^2 \langle \bar{d} d \rangle \langle g_s^2 G^2 \rangle \dfrac{m_s (5 t^2+2 t+5)}{3^2\times 2^{19}M^2 s_0 \pi^6}\left[ 2 s_0 +3(M^2+s_0 ln( \dfrac{s}{\Lambda^2}) )e^{-\dfrac{s_0}{M^2}} \right] 
\nonumber \\
&+&
\langle \bar{d} d \rangle^3  \langle g_s^2 G^2 \rangle \dfrac{g_s^2 m_s}{3^{6}\times 2^{12}M^2 s_0 \pi^4}\left[s_0 (13 t^2+ 10 t+13)+3(5 t^2+2 t+5) (M^2+s_0 ln( \dfrac{s}{\Lambda^2}) )e^{-\dfrac{s_0}{M^2}}\right] 
\nonumber \\
&+&
\langle \bar{d} d \rangle^2 \langle \bar{s} s \rangle \langle g_s^2 G^2 \rangle \dfrac{g_s^2 m_s (t^2+1)}{3^7\times 2^{13} M^2 \pi^4}+
\langle \bar{d} d \rangle^2 \langle \bar{u} u \rangle \langle g_s^2 G^2 \rangle \dfrac{g_s^2 m_s (t^2-1)}{3^2\times 2^{9} M^2 s_0 \pi^2}\left[ 2 s_0+ 3(M^2+s_0 ln( \dfrac{s}{\Lambda^2})) e^{-\dfrac{s_0}{M^2}}\right] 
\nonumber \\
&+&
m_0^4\langle \bar{d} d \rangle^2 \langle \bar{s} s \rangle \dfrac{m_s (t^2-1)}{2^8 M^2 s_0 \pi^2} \left[ -3s_0+4\left( M^2+s_0 ln( \dfrac{s}{\Lambda^2})\right) e^{-\dfrac{s_0}{M^2}}\right] - m_0^4\langle \bar{d} d \rangle \langle \bar{s} s \rangle \langle \bar{u} u \rangle \dfrac{5 m_s(t^2-1)}{3\times 2^{11} M^2 \pi^2}
\nonumber \\
&+&
\langle \bar{d} d \rangle \langle \bar{u} u \rangle^2 \langle g_s^2 G^2 \rangle \dfrac{m_s}{3^6\times 2^{14} M^6 s_0 \pi^4} \left[ s_0(t-1)^2 (8M^4(g_s^2+108 \pi^2)-81m_0^4 \pi^2) 
\right. 
\nonumber \\
&+&
\left. 
24 M^4 (54 \pi^2 (t-1)^2+g_s^2(5t^2+2t+5))(M^2+s_0 ln( \dfrac{s}{\Lambda^2}))e^{-\dfrac{s_0}{M^2}}\right]
\nonumber \\
&-&
m_0^4  \langle \bar{d} d \rangle \langle \bar{u} u \rangle^2  \dfrac{m_s (t-1)^2}{3\times 2^{11} M^2 s_0 \pi^2} \left[ s_0 -4 (M^2+s_0 ln( \dfrac{s}{\Lambda^2}))e^{-\dfrac{s_0}{M^2}}\right] 
\nonumber \\
&+&\langle \bar{u} u \rangle^2 \langle \bar{s} s \rangle \langle g_s^2 G^2 \rangle \dfrac{g_s^2 m_s (3 t^2 + 2 t +3)}{3^5\times 2^{11}M^2 \pi^4}-m_0^4  \langle \bar{s} s \rangle \langle \bar{u} u \rangle^2  \dfrac{5 m_s (t-1)^2}{3^2\times 2^{11} M^2 \pi^2 },
\end{eqnarray}

\begin{eqnarray}
\tilde{\Pi}^{Dim14}&=&-\langle \bar{d} d \rangle ^2 \langle g_s^2 G^2 \rangle^2 \dfrac{g_s^2 (t+1)^2}{3^6\times 2^{19} M^2 \pi^6}+m_0^4\langle \bar{d} d \rangle \langle \bar{s} s \rangle \langle g_s^2 G^2 \rangle\dfrac{5 t^2 + 2 t + 5}{3\times 2^{16} M^2 \pi^4}-m_0^2\langle \bar{d} d \rangle^3  \langle \bar{s} s \rangle\dfrac{g_s^2(5 t^2 + 2 t+5)}{3^5 \times 2^{8} M^2 \pi^2}
\nonumber \\
&+&
m_0^4\langle \bar{d} d \rangle\langle \bar{u} u \rangle\langle g_s^2 G^2 \rangle \dfrac{t^2-1}{2^{14} M^2 \pi^4}-m_0^2 \langle \bar{d} d \rangle^3 \langle \bar{u} u \rangle \dfrac{g_s^2 (t^2-1)}{3^4\times 2^{7} M^2 \pi^2}-m_0^2 \langle \bar{d} d \rangle^2 \langle \bar{s} s \rangle \langle \bar{u} u \rangle \dfrac{t^2-1}{3\times 2^4 M^2}
\nonumber \\
&-&
m_0^2 \langle \bar{s} s \rangle^2 \langle \bar{u} u \rangle \langle \bar{d} d \rangle \dfrac{g_s^2 (t^2-1)}{3^4\times 2^{7} M^2 \pi^2}+ \langle g_s^2 G^2 \rangle^2 \langle \bar{u} u \rangle ^2 \dfrac{g_s^2 (29 t^2 +50 t+29)}{3^6 \times 2^{19}M^2 \pi^6}-m_0^2 \langle \bar{d} d \rangle^2 \langle \bar{u} u \rangle^2 \dfrac{g_s^2(t-1)^2}{3^6\times 2^{9} M^2 \pi^2}
\nonumber \\
&-&m_0^2 \langle \bar{u} u \rangle^2 \langle \bar{d} d \rangle \langle \bar{s} s \rangle \dfrac{\left[ 81 \pi^2 (t-1)^2+g_s^2 (5 t^2 +2 t +5)\right] }{3^5\times 2^{7} M^2 \pi^2}-m_0^2 \langle \bar{u} u \rangle^2 \langle \bar{s} s \rangle^2 \dfrac{g_s^2 (t-1)^2}{3^5 \times 2^{9} M^2 \pi^2}
\nonumber \\
&-&
m_0^2 \langle \bar{u} u \rangle^3 \langle \bar{d} d \rangle \dfrac{g_s^2 (t^2-1)}{3^4 \times 2^{7} M^2 \pi^2},
\nonumber \\
\tilde{\Pi}^{Dim15}&=&-m_0^2 \langle \bar{d} d \rangle^3 \langle g_s^2 G^2 \rangle \dfrac{m_s (119 t^2+46 t+119)}{3^6\times 2^{17}M^4 \pi^4} -\left( m_0^2 \langle \bar{d} d \rangle^2 \langle \bar{u} u \rangle \langle g_s^2 G^2 \rangle + m_0^6  \langle \bar{d} d \rangle^2  \langle \bar{u} u \rangle\right)  \dfrac{m_s (t^2-1)}{2^{11}\times M^4 \pi^2}
\nonumber \\
&-&m_0^2 \langle g_s^2 G^2 \rangle  \langle \bar{d} d \rangle  \langle \bar{u} u \rangle^2  \dfrac{m_s}{3^6 \times 2^{16} M^4 \pi^4} \left[ 2592 \pi^2 (t-1)^2+g_s^2 (123 t^2 + 50 t +123)\right] 
\nonumber \\
&-& m_0^6 \langle \bar{d} d \rangle \langle \bar{u} u \rangle^2 \dfrac{m_s (t-1)^2}{3\times 2^{12} M^4 \pi^2},
\nonumber \\
\tilde{\Pi}^{Dim16}&=&0,
\nonumber \\
\tilde{\Pi}^{Dim17}&=&  \langle g_s^2 G^2 \rangle^2\langle \bar{d} d \rangle^3 \dfrac{g_s^2 m_s (t+1)^2}{3^6\times 2^{15} M^6 \pi^4}-m_0^4 \langle g_s^2 G^2 \rangle \langle \bar{d} d \rangle^2 \langle \bar{u} u \rangle \dfrac{m_s(t^2-1)}{2^{13} M^6 \pi^2}
\nonumber \\
&+&m_0^2 \langle \bar{d} d \rangle^3 \langle \bar{u} u \rangle^2 \dfrac{g_s^2 m_s}{3^9\times 2^{7}M^6 \pi^2} \left[54\pi^2(t-1)^2+g_s^2(5 t^2 +2 t +5) \right] 
\nonumber \\
&-&
\langle g_s^2 G^2 \rangle^2 \langle \bar{u} u \rangle^2 \langle \bar{d} d \rangle \dfrac{g_s^2 m_s (3 t^2 +2 t+ 3)}{3^6\times 2^{14}M^6 \pi^4}+
m_0^2 \langle \bar{u} u \rangle^3 \langle \bar{d} d \rangle^2 \dfrac{g_s^2 m_s (t^2-1)}{3^5\times 2^5 M^6}
\nonumber \\
&-&m_0^4 \langle g_s^2 G^2 \rangle\langle \bar{u} u \rangle^2 \langle \bar{d} d \rangle \dfrac{m_s (t-1)^2}{3^2\times 2^{14}M^{6}\pi^2}-m_0^4 \langle g_s^2 G^2 \rangle\langle \bar{u} u \rangle \langle \bar{d} d \rangle^2 \dfrac{m_s (t^2-1)}{ 2^{13}M^{6}\pi^2}
\nonumber \\
&+& m_0^2 \langle \bar{u} u \rangle^4 \langle \bar{d} d \rangle \dfrac{g_s^2 m_s (5 t^2 +2 t+ 5)}{3^9\times 2^8 M^6 \pi^2},
\nonumber \\
\tilde{\Pi}^{Dim18}&=&0,
\nonumber \\
\tilde{\Pi}^{Dim19}&=&m_0^2 \langle g_s^2 G^2 \rangle^2 \langle \bar{d} d \rangle ^3 \dfrac{g_s^2 m_s (t+1)^2}{3^6\times 2^{16}M^8 \pi^4}-m_0^6 \langle g_s^2 G^2 \rangle \langle \bar{d} d \rangle ^2 \langle \bar{u} u \rangle \dfrac{m_s (t^2-1)}{3\times 2^{14}M^8 \pi^2}
\nonumber \\
&+&  \langle g_s^2 G^2 \rangle \langle \bar{d} d \rangle ^3 \langle \bar{u} u \rangle ^2 \dfrac{g_s^2 m_s }{3^{10}\times 2^{8} M^8 \pi^2} \left[ 54 \pi^2 (t-1)^2 +g_s^2 (5 t^2 +2 t+ 5) \right] 
\nonumber \\
&+& m_0^4 \langle \bar{d} d \rangle ^3 \langle \bar{u} u \rangle ^2  \dfrac{g_s^2 m_s (t-1)^2}{3^6\times 2^{6}M^8}-m_0^2 \langle g_s^2 G^2 \rangle^2 \langle \bar{u} u \rangle ^2 \langle \bar{d} d \rangle \dfrac{g_s^2 m_s (3 t^2 +2 t+ 3)}{3^6\times 2^{15}M^8 \pi^4}
\nonumber \\
&+&\langle g_s^2 G^2 \rangle \langle \bar{d} d \rangle ^2 \langle \bar{u} u \rangle ^3  \dfrac{g_s^2 m_s (t^2-1)}{3^6 \times 2^{6} M^8}+ m_0^4 \langle \bar{d} d \rangle ^2  \langle \bar{u} u \rangle ^3 \dfrac{g_s^2 m_s (t^2-1)}{3^5 \times 2^5 M^8}
\nonumber \\
&+&\langle g_s^2 G^2 \rangle \langle \bar{d} d \rangle \langle \bar{u} u \rangle ^4 \dfrac{g_s^4 m_s (4 M^2+3 m_0^2)(5 t^2 +2 t+ 5)}{3^{10}\times 2^{11}M^{10}\pi^2},
\nonumber \\
\tilde{\Pi}^{Dim20}&=&0,
\nonumber \\
\tilde{\Pi}^{Dim21}&=& m_0^2 \langle g_s^2 G^2 \rangle \langle \bar{d} d \rangle ^3\langle \bar{u} u \rangle ^2 \dfrac{g_s^2 m_s}{3^9\times 2^{10}M^{10}\pi^2} +m_0^6\langle \bar{d} d \rangle ^3\langle \bar{u} u \rangle ^2  \dfrac{g_s^2 m_s (t-1)^2}{3^5\times 2^9 M^{10}}
\nonumber \\
&+& m_0^2  \langle g_s^2 G^2 \rangle \langle \bar{d} d \rangle\langle \bar{u} u \rangle ^4 \dfrac{g_s^4 m_s (5 t^2 +2 t + 5)}{3^9\times 2^{11}M^{10}\pi^2} +m_0^2 \langle g_s^2 G^2 \rangle\langle \bar{d} d \rangle^2 \langle \bar{u} u \rangle^3 \dfrac{g_s^2 m_s (t^2-1)}{3^{4}\times 2^{8}M^{10}}
\nonumber \\
&+& m_0^6 \langle \bar{d} d \rangle^2 \langle \bar{u} u \rangle^3 \dfrac{g_s^2 m_s (t^2-1)}{3^4\times 2^{8}M^{10}}
,
\nonumber \\
\tilde{\Pi}^{Dim22}&=&0,
\nonumber \\
\tilde{\Pi}^{Dim23}&=& m_0^4  \langle g_s^2 G^2 \rangle \langle \bar{d} d \rangle^3  \langle \bar{u} u \rangle^2 \dfrac{g_s^2 m_s (t-1)^2}{3^5\times 2^{9}M^{12}}+m_0^4  \langle g_s^2 G^2 \rangle \langle \bar{d} d \rangle^2  \langle \bar{u} u \rangle^3 \dfrac{g_s^2 m_s (t^2-1)}{3^4\times 2^{8} M^{12}}
\nonumber \\
&-&m_0^2 \langle \bar{d} d \rangle^3\langle \bar{u} u \rangle^4\dfrac{g_s^6 m_s (5 t^2 +2 t+ 5)}{3^{13}\times 2^{3}M^{12}},
\nonumber \\
\tilde{\Pi}^{Dim24}&=&0,
\end{eqnarray}

\begin{eqnarray}
\tilde{\Pi}^{Dim25}&=&m_0^6  \langle g_s^2 G^2 \rangle \langle \bar{d} d \rangle^3  \langle \bar{u} u \rangle^2 \dfrac{5 g_s^2 m_s (t-1)^2}{3^6\times 2^{11}M^{14}}+ m_0^6  \langle g_s^2 G^2 \rangle \langle \bar{d} d \rangle^2  \langle \bar{u} u \rangle^3 \dfrac{5 g_s^2 m_s (t^2-1)}{3^5\times 2^{10}M^{14}}
\nonumber \\
&-& \langle g_s^2 G^2 \rangle \langle \bar{d} d \rangle^3  \langle \bar{u} u \rangle^4 \dfrac{5 g_s^2 m_s (2 M^2+3 m_0^2)(5 t^2 +2 t+5)}{3^{14}\times 2^{6}M^{16}},
\nonumber \\
\tilde{\Pi}^{Dim26}&=&0,
\nonumber \\
\tilde{\Pi}^{Dim27}&=&-m_0^2 \langle g_s^2 G^2 \rangle\langle \bar{d} d \rangle^3 \langle \bar{u} u \rangle^4  \dfrac{g_s^6 m_s (5 t^2 +2 t+5 )}{3^{13}\times 2^{6}M^{16}}.
\end{eqnarray}
In the above expressions, $t=\tan \theta$, Euler constant $\gamma_E \simeq 0.577$,  and  $\Lambda=0.5$~GeV is the QCD scale parameter. The masses of  the light $u$ and $d$ quarks are taken as zero for the sake of brevity in presenting the results.

\section*{ACKNOWLEDGEMENTS}
K. Azizi is thankful to Iran Science Elites Federation (Saramadan)
for the partial  financial support provided under the grant number ISEF/M/401385.



\end{document}